\documentclass{ws-procs975x65}

\newcommand{\bea}{\begin{eqnarray}}
\newcommand{\eea}{\end{eqnarray}}

\newcommand{\be}{\begin{eqnarray}}
\newcommand{\ee}{\end{eqnarray}}
\def\lsim{\mathrel{\rlap{\lower3pt\hbox{\hskip1pt$\sim$}}
     \raise1pt\hbox{$<$}}} 
\def\gsim{\mathrel{\rlap{\lower3pt\hbox{\hskip1pt$\sim$}}
     \raise1pt\hbox{$>$}}} 

\def\prl{Phys. Rev. Lett.}
\def\np{Nucl. Phys.}
\def\pr{Phys. Rev.}
\def\pl{Phys. Lett.}
\def\la{\langle}\def\ra{\rangle}
\def\del{\partial}
\def\calL{\cal L}

\newcommand\Tr{\mathrm{Tr}\,}

\def\bi{\bibitem}

\let\emph=\relax

\let\mathbf=\boldsymbol
\def\beginABC{\begin{subequations}}
\def\endABC{\end{subequations}}

\begin{document}

\title{Half-Skyrmion Hadronic Matter at High Density}

\author{Hyun Kyu Lee$^1$ and Mannque Rho$^{1,2}$}

\address{$^1$ Department of Physics, Hanyang University, Seoul 133-791, Korea\\
$^2$ Institut de Physique Th\'eorique, CEA Saclay, 91191 Gif-sur-Yvette, France}

\begin{abstract}

The hadronic matter described as a skyrmion matter embedded in an FCC crystal is found to turn into a half-skyrmion matter with vanishing (in the chiral limit) quark condensate and {\em non-vanishing} pion decay constant at a density $n_{1/2S}$ lower than or at the critical density $n_c^{\chi SR}$ at which hadronic matter changes over to a chiral symmetry restored phase with deconfined quarks. When hidden local gauge fields and dilaton scalars -- one ``soft" and one ``hard" -- are incorporated,  this phase is characterized by $a=1$, $f_\pi\neq 0$ with the hidden gauge coupling $g\neq 0$ but $\ll 1$. While chiral symmetry is restored in this region in the sense that $\la\bar{q}q\ra=0$, quarks are still confined in massive hadrons and massless pions. This phase seems to correspond to the ``quarkyonic phase" predicted in large $N_c$ QCD. It also represents the ``hadronic freedom" regime relevant to kaon condensation at compact-star density. As $g\rightarrow 0$ (in the chiral limit), the symmetry ``swells" -- as an emergent symmetry due to medium -- to $SU(N_f)^4$ as proposed by Georgi for the ``vector limit." The fractionization of skyrmion matter into half-skyrmion matter is analogous to what appears to happen in condensed matter in (2+1) dimensions where  half-skyrmions or ``merons" enter as relevant degrees of freedom at the interface. Finally the transition from baryonic matter to color-flavor-locked quark matter can be bridged by a half-skyrmion matter.

\end{abstract}

\body

\section{Introduction}
Hadronic matter at high density is presently poorly understood and the issue of the equation of state (EOS) in the density regime appropriate for the interior of compact stars remains a wide open problem. Unlike at high temperature where lattice QCD backed by relativistic heavy ion experiments is providing valuable insight into hot medium, the situation is drastically different for cold hadronic matter at a density a few times that of the ordinary nuclear matter relevant for compact stars.  While asymptotic freedom should allow perturbative QCD to make well-controlled predictions at superhigh densities, at the density regime relevant for compact stars, there are presently neither reliable theoretical tools nor experimental guides available to make clear-cut statements. The lattice method, so helpful in high-T matter, is hampered by the sign problem and cannot as yet handle the relevant  density regime.

What is generally accepted at the moment is that effective field theories formulated in terms of hadronic variables, guided by a wealth of experimental data, can accurately describe baryonic matter up to nuclear matter density $n_0\approx 0.16$ fm$^{-3}$ and perturbative QCD unambiguously predicts that color superconductivity should take place in the form of color flavor locking (CFL) at some asymptotically high density $n_{CFL}$~\cite{cfl-rev}. In between, say, $n_0\leq n \leq n_{CFL}$, presently available in the literature  are a large variety of model calculations which however have not been checked by first-principle theories or by experiments. The model calculations so far performed paint a complex landscape of phases from  $n_0$ to  $n_{CFL}$, starting with kaon condensation at $n_c^K \sim  3n_0$~\cite{BLR07}, followed by a plethora of color superconducting quark matter with or without color flavor locking near and above  the chiral restoration $n_c^{\chi SR}$ and ultimately CFL with or without kaon condensation. It is unclear which of the multitude of the phases could be realized and how they would manifest themselves in nature.

In this note, we would like to zero in on the vicinity of the chiral restoration
density denoted $n_c^{\chi SR}$ at which both the quark condensate $\la\bar{q}q\ra$ and the physical pion decay constant $f_\pi$ go to zero in the chiral limit and explore a hitherto unsuspected novel
phenomenon that could take place very near $n_c^{\chi SR}$. This can be efficiently done by putting skyrmions in a crystal lattice.~\footnote{No formulation for the phenomenon described in this note is available in continuum but we expect the topological structure will remain intact in the continuum limit.}
While the skyrmion structure has been extensively studied in hadronic physics as a description of a baryon in QCD at large $N_c$, one expects it to equally provide a powerful approach to many-body systems: A skyrmion with winding number $B$ is to encode entire strong interactions of QCD at large $N_c$ for systems with $B$ baryons. Thus the skyrmion description has the potential to provide a {\it unified approach} to baryonic dynamics, not only that of elementary baryons but also the structure of complex nuclei as well as infinite matter at any density, both below and above the deconfinement point. Perhaps academic but theoretically fascinating  is the possibility that the CFL phase can also be described as a skyrmion matter of different form --  to be referred to as ``superqualiton" matter. Thus the transition from normal matter to CFL matter can be considered as a skyrmion-superqualiton  transition, with a half-skyrmion phase as the border between the two~\cite{MR-halfskyrmion}.

Treating dense nuclear matter in terms of skyrmion
matter, we will argue that at a density denoted $n_{1/2S}$ lying at $\lsim n_c^{\chi SR}$, a skyrmion in
dense matter fractionizes into two half skyrmions with chiral
$SU(N_f)\times SU(N_f)$ symmetry restored but with a {\em non-vanishing}
pion decay constant $f_\pi\neq 0$.  Phrased in terms of hidden local symmetry Lagrangian where the lowest-lying vector mesons $\rho$ and $\omega$ are introduced in addition to the Nambu-Goldstone pions, the chiral symmetry is restored in the sense that $f_\pi=f_\sigma\neq 0$ where
\be
\la\pi^b(q)|A_\mu^a|0\ra=iq_\mu\delta^{ab}f_\pi, \ \ \la\sigma^b(q)|V_\mu^a|0\ra=iq_\mu\delta^{ab}f_\sigma
\ee
but with $\la\bar{q}q\ra=0$, where $\sigma^b$ is the Nambu-Goldstone boson to be higgsed to become the longitudinal component of the $\rho$ meson. Here chiral symmetry is restored but the phase transition scenario differs from the standard Nambu-Goldstone-to-Wigner Weyl transition in that $f_\pi\neq 0$ in this phase. To distinguish this phase from the usual chiral symmetry restored phase with $\la\bar{q}q\ra\sim f_\pi=0$, we shall refer to it as ``1/2-skyrmion phase." This phase could be identified with the ``vector symmetry" of Georgi~\cite{georgi}.~\footnote{It was argued by Harada and Yamawaki~\cite{HY:PR}that the vector limit with $g=0$ and $f_\pi=f_\sigma\neq 0$ does not satisfy the axial Ward identity and that it is the limit $g=f_\pi=0$, called ``vector manifestation," that does. A comment will be made on this point later.}

It should be noted that in this half-skyrmion phase, quarks are still confined although chiral symmetry is restored. Thus it resembles the ``quarkyonic phase" predicted~\cite{quarkyonic} in the large $N_c$ limit of QCD characterized by the order parameter $B_0$ in which chiral symmetry is restored but the quarks are confined. In this phase baryons in which the quarks are confined are massive, so cannot enter in the 't Hooft anomaly condition~\footnote{The anomaly matching condition states that a composite particle has to reproduce exactly the anomlay present in the fundamental theory, that is to say that the fundamental anomaly and  the anomalies in the composite theory must match. For this matching to be satisfied by the composite system, there must exist massless excitations. It has been shown that this matching condition holds in the presence of chemical potential~\cite{hsu}.}. The 't Hooft anomaly matching could be assured in the 1/2-skyrmion phase by the massless pion which is present.

We suggest that the phase $n_{1/2S}\lsim n\lsim n_c^{\chi SR}$ can also be identified with the
``hadronic freedom" regime and $n_{1/2S}$ as the ``flash density" $n_{flash}$~\cite{BLR07}, both
of which play an important role in describing dense matter near and
just below the chiral transition point.~\footnote{The corresponding temperature in hot medium is called ``flash temperature." More on this below in connection with heavy ion collisions~\cite{BHHRS}.} There is also a tantalizing analogy between the half-skyrmion phase present in dense matter and the meron phases in (2+1) dimensions encountered in condensed matter.

\section{Vector Mesons and Dilatons in Skyrmion Matter}
Up to date, most of the works done on skyrmions relied on the Skyrme Lagrangian that contains the  current algebra term and the Skyrme term~\footnote{We reserve $f_\pi$ for the physical pion decay constant while $F_\pi$ stands for a parameter in the Lagrangian. In the mean field approximation used below, they are equivalent.}, viz,
\be
{\cal L}= \frac{F_\pi^2}{4} {\Tr} (\del_\mu U\del^\mu U^\dagger)+\frac{1}{32e^2}{\Tr} [U^\dagger\del_\mu U, U^\dagger\del^\nu U]^2
 \ee
implemented with mass terms. But there are compelling reasons to believe that other degrees of freedom than the pions are essential for reliably describing systems with $B > 1$. It seems certain that both vector and scalar excitations are essential. It has in fact been argued since sometime that vector mesons must figure in the topological structure of elementary baryon as well as baryonic matter~\cite{MR-Yukawa}~\footnote{A glaring defect of the skyrmion with pion fields but with no other fields (such as the vector mesons $\rho$, $\omega$ etc) is that when applied to nuclei, the parameters needed to even approximately fit
nature are totally unnatural. For instance, the parameter $f_\pi$ is
much too small compared with the physical value $f_\pi\approx 93$
MeV -- this is so even for a single nucleon -- and the pion mass
parameter $m_\pi$ is much too large compared with its free-space
value $m_\pi\approx 140$ MeV. See e.g. \cite{battye-sutcliffe}.
When the parameters are taken to be close to their physical values,
the resulting structure at the mean field level of complex nuclei,
e.g., shape, comes out to be completely different from what is known
in nature.}. Indeed the recent development in holographic dual QCD (hQCD)~\cite{SSmodel} indicates that not just the lowest vector mesons but the infinite tower of vector mesons encapsulated in five-dimensional (5D) Yang-Mills Lagrangian can drastically modify the structure of baryons arising as instantons~\cite{HRYY,SS,Kim-Zahed}. This suggests that dense matter described with a hidden local symmetric Lagrangian with the infinite tower would be drastically different from the picture given by the pion-only skyrmion description. This point will be addressed below. Furthermore it has become evident that certain scalar degrees of freedom associated with the trace anomaly of QCD could also figure crucially~\cite{PRV08,LR09}. This development came about in implementing broken scale invariance in the skyrmion structure of dense matter built in the presence of vector mesons. A remarkable structure arises in the presence of the $\omega$ meson and two scalar mesons corresponding to the dilatons of spontaneously broken scale invariance as we will describe.
\subsection{Dilatons in hidden local symmetry}
To start with, let us describe the Lagrangian with which we will develop our arguments. To bring out the notion that hidden local symmetry in low-energy dynamics is quite generic, it is instructive to see how hidden local fields ``emerge" naturally from a low-energy theory~\cite{emergent}. As will be noted, the same structure can be obtained top-down from string theory.

The idea is simply that the chiral field $U=e^{2i\pi/F_\pi}$ -- which represent the coordinates for the symmetry $SU(N_f)_L\times SU(N_f)_R/SU(N_f)_{L+R}$ -- can be written in terms of the left and right coset-space coordinates as
 \be
U=\xi_L^\dagger\xi_R\label{Up}
 \ee
with transformation under $SU(N_f)_L\times SU(N_f)_R$ as
$\xi_L\rightarrow \xi_L L^\dagger$ and $\xi_R\rightarrow \xi_R R^\dagger$ with
$L(R)\in SU(N_f)_{L(R)}$.  Now the redundancy that is hidden, namely, the invariance under the $local$
transformation
 \be
\xi_{L,R}\rightarrow h(x)\xi_{L,R}
 \ee
where $h(x)\in SU(N_f)_{V=L+R}$ can be elevated to a local gauge invariance~\cite{HY:PR} with the corresponding gauge field
$V_\mu\in SU(N_f)_{V}$ that transforms
 \be
V_\mu\rightarrow h(x)(V_\mu +i\del_\mu)h^\dagger (x).
 \ee
The resulting  HLS Lagrangian given in terms of the covariant derivative $D_\mu$ takes the form \cite{georgi} (with $V_\mu=g\rho_\mu$):
 \be
{\calL}= \frac{F_\pi^2}{4}{\Tr}\{|D_\mu\xi_L|^2 + |D_\mu\xi_R|^2
 + \gamma |D_\mu U|^2\} \ - \frac{1}{2} \,
\mbox{Tr} \left[ \rho_{\mu\nu} \rho^{\mu\nu} \right] +\cdots \label{hls1}
 \ee
where the ellipsis stands for higher derivative and other higher dimension terms including the gauged Skyrme term. If one parameterizes $\xi_{L,R}=e^{i\sigma/F_\sigma}e^{\mp i\pi/F_\pi}$, gauge-fixing with $\sigma=0$ corresponds to unitary gauge, giving  the usual gauged nonlinear sigma model with a mass term for the gauge field. Clearly one can extend such a construction to an infinite tower of vector mesons spread in energy in the fifth dimension. Such a construction has been made and led to the so-called ``dimensionally deconstructed QCD" encapsulated in a 5D Yang-Mills theory~\cite{son-stephanov}. The latter is essentially equivalent in form to the 5D Yang-Mills theory of holographic dual QCD that comes from string theory~\cite{SSmodel}. This infinite-tower HLS theory will be denoted as HLS$_\infty$. As noted by Harada, Matsuzaki and Yamawaki~\cite{HMY}, the Lagrangian (\ref{hls1}) -- denoted in an obvious notation as HLS$_1$ -- can be thought of as a truncated version of HLS$_\infty$ where {\em all} other than the lowest vector mesons $\rho$ and $\omega$ are integrated out.

For studying the properties of dense hadronic matter, the scaling behavior of the effective Lagrangian is crucial. In fact the early description of how hadron properties change in hot/dense medium was anchored on the role played by the scalar dilaton associated with the trace anomaly of QCD~\cite{BR91}. It was clear then that the spontaneous breaking of chiral symmetry which leads to the generation of hadron masses and the explicit breaking of scale invariance by the quantum anomaly in QCD, which brings a length scale, must be connected. How to introduce scalar degrees of freedom to the HLS Lagrangian (\ref{hls1}) is, however, not so straightforward since both chiral symmetry breaking ($\chi$SB) and confinement are intricately involved. In \cite{LR09}, this problem was solved by introducing two dilatons, one ``soft" and the other ``hard," with the soft dilaton $\chi_s$ intervening in $\chi$SB and the hard dilaton $\chi_h$ intervening in confinement-deconfinement. By integrating out the latter to focus on the chiral symmetry properties of hadrons, a suitable HLS$_1$ Lagrangian was obtained in \cite{LR09}. Written in unitary gauge and with some harmless simplifications, it takes the form (including the pion mass term) for two light flavors (up and down)~\footnote{For flavor number $N_f < 3$, the well-known 5D topological Wess-Zumino term is absent. However in the presence of vector mesons as in hidden local symmetry formulation, there are in general four terms -- that we shall call ``hWZ terms" -- in the anomalous parity sector that satisfy homogenous anomaly equation. It turns out that if one requires vector dominance in photon-induced processes involving the hWZ terms and use the equation of motion for a heavy  $\rho$ field, then the hWZ terms can be reduced to one term in the regime we are concerned with as given in this formula.}:
\bea
{\cal L}={\cal L}_{\chi_s}+ {\cal L}_{hWZ}\label{lagtot}
\eea
where
\bea
{\cal L}_{\chi_s} &=& \frac{F_\pi^2}{4}\kappa^2
\mbox{Tr}(\partial_\mu U^\dagger \partial^\mu U) + \kappa^3 v^3 \mbox{Tr}M(U+U^\dagger)
\nonumber\\
&&
-\frac{F_\pi^2}{4} a\kappa^2 \mbox{Tr}[\ell_\mu + r_\mu + i(g/2)
( \vec{\tau}\cdot\vec{\rho}_\mu + \omega_\mu)]^2\nonumber\\
&& -\textstyle \frac{1}{4} \displaystyle
\vec{\rho}_{\mu\nu} \cdot \vec{\rho}^{\mu\nu}
-\textstyle \frac{1}{4}  \omega_{\mu\nu} \omega^{\mu\nu}
+\frac 12 \del_\mu\chi_s\del^\mu\chi_s + V(\chi_s)\label{lags}\\
{\cal L}_{hWZ} &=& \textstyle\frac{3}{2} g \kappa^3 \omega_\mu B^\mu\label{fwzterm}
\eea
where $\kappa=\chi_{s}/f_{\chi_{s}}$ with $f_{\chi_s}=\la 0|\chi_s|0\ra$ and
\bea
B^\mu =  \frac{1}{24\pi^2} \varepsilon^{\mu\nu\alpha\beta}
\mbox{Tr}(U^\dagger\partial_\nu U U^\dagger\partial_\alpha U
U^\dagger\partial_\beta U)
\eea is the baryon current and
\bea
V(\chi_{s})=B\chi_{s}^4{\rm ln}\frac{\chi_{s}}{f_{\chi_{s}}e^{1/4}}\label{potterm}
\eea
is the dilaton potential.

\subsection{The 1/2-skyrmion matter}

For understanding a generic feature of dense skyrmion matter, it is illuminating to first consider the Skyrme Lagrangian coupled to the dilaton $\chi_s$ which is gotten from (\ref{lagtot}) by setting $\rho_\mu=\omega_\mu=0$ and putting a quartic Skyrme term to assure the topological stability. There have been a series of works on dense matter treated with this Skyrme-dilaton Lagrangian~\cite{Skyrme-matter} on which we will base our beginning arguments.

In \cite{Skyrme-matter}, following the seminal work of Klebanov~\cite{klebanov}, density effect is simulated by putting skyrmions in a crystal and squeezing the crystal. In (3+1) dimensions, it is found to be energetically favorable to arrange the skyrmions as a face-centered cubic crystal (FCC) lattice~\cite{FCC}. One should however recognize that there is no proof that this is indeed the absolute minimal configuration. There may be other configurations that are more favorable. Indeed, it has been recently shown that in baby-skyrmion systems~\cite{karliner}, of all possible crystalline structures, it is  the hexagonal, not the cubic, that  gives the minimal energy.   This caveat notwithstanding,  we will base our discussions on the FCC crystalline structure.   We will say more on this below, in particular concerning certain qualitative features that could be different for different crystalline structures.

Briefly, what is done in \cite{Skyrme-matter} is as follows. The crystal configuration made up of skyrmions has each FCC lattice site occupied by a single skyrmion centered with $U_0=-1$ with each nearest neighbor pair relatively rotated in isospin space  by $\pi$ with respect to the line joining the pair. In order to have the Skyrme Lagrangian possess the correct scaling under scale change of the crystalline, the dilaton scalar $\chi$ associated with the trace anomaly of QCD has to be implemented as suggested in \cite{BR91}. The energy density of the lattice skyrmions is then given by~\cite{Skyrme-matter}
\bea
\epsilon &=& \frac{1}{4} \int_{Box} d^3x \left\{ \frac{f^2_\pi}{4} \left(
\frac{\chi^*}{f_\chi} \right)^2 \mbox{Tr}(\partial_i U_0^\dagger
\partial_i U_0) + \cdots
\right.
\nonumber \\
&& \left. \hskip 1em
+ \frac12 \partial_i \chi^* \partial_i \chi^* + V(\chi) \right\}
\label{EoverB}\end{eqnarray}
where $f_\pi$ is the physical pion decay constant (which is equal to the parametric constant $F_\pi$ at the tree order). We have dropped the subscript ``s" since we are dealing only with $\chi_s$. Here, the ellipsis stands for the familiar Skyrme quartic term and quark mass terms which need not be explicited, the subscript `box' denotes that the integration is over a single FCC box and the factor 1/4 in front appears because the box contains baryon number four. The asterisk ``*" denotes the mean field (a density-dependent object for $n\neq 0$), $f_\chi$ is the $\chi$ decay constant and $V(\chi)$ is the dilaton potential for $\chi_s$~\cite{BHLR06,LR09}. The field $\chi$ is coupled to the chiral field $U$, so the mean field $ \chi^*=\la \chi\ra_n$ (for a given density $n$) scales with the background provided by the crystal configuration. The minimization of this energy density with respect to the coefficients of the Fourier expansion of the (mean) fields taken as variational parameters reveals that at some minimum size of the box corresponding to a density, say, $n_{1/2S}$  of the matter, there is a phase transition from the FCC crystal configuration of skyrmions into a body-centered cubic crystal (BCC) configuration of half skyrmions as predicted on symmetry grounds~\cite{goldhaber-manton,FCC}. We should point out  two aspects here that characterize the transition. One is that what is involved here is a topology change, also observed  in (2+1) dimensions. Therefore it has the possibility of being stable against quantum fluctuations. The other is that in terms of the mean chiral field $U_0(x)=\sigma (x) +i\tau\cdot\pi$, the expectation value $\la\sigma\ra \propto \la\bar{q}q\ra$ is zero at $n_{1/2S}$, so the transition is indeed a chiral restoration phase transition.

The result of the calculation in \cite{Skyrme-matter} uncovers several striking features in skyrmions at dense matter. The most prominent among them is that while chiral symmetry is restored at $n_{1/2S}$, the pion decay constant given by $f_\pi^*/f_\pi\propto \la\chi\ra_{n\geq n_p}/f_\chi\neq 0$. In terms of the chiral order parameter $\la\bar{q}q\ra\sim \la c{\Tr} (U+U^\dagger)\ra$, the 1/2-skyrmion phase has $\la{\Tr}(U+U^\dagger)\ra=0$ but $c\sim f_\pi^* \neq 0$. A similar property has been proposed for high temperature and identified with a ``pseudogap phase" in analogy to high T superconductivity~\cite{zarembo}. Note that this phase is distinct from the standard $\chi$SB phase where the pion decay constant is directly proportional to the quark condensate. Although the connection is not clear, the phase between $n_p$ and the density denoted $n_c^{\chi SR}$ at which $f_\pi^*=0$ is called ``pseudogap phase." The density range of the pseudogap phase depends on the mass of the scalar $\chi$. As will be seen, the range can be shrunk to a point for certain value of the mass for the dilaton but there is always some region in which the 1/2-skyrmion phase is the lowest energy state.
\subsection{The effect of the $\omega$ meson}
In the presence of vector-meson fields, particularly the $\omega$ meson field, the phase structure is dramatically different from the one without vector mesons. With vector mesons, in particular, with the $\omega$, the dilaton $\chi$ plays a crucial role. There is a close interplay between the $\omega$ which supplies repulsion between skyrmions (nucleons) and the dilaton which provides attraction that leads to the binding in nuclei~\footnote{It is worth mentioning here that the scalar ``$\sigma$" in Walecka mean-field theory corresponds to this dilaton, not to the fourth component of the chiral four-vector in linear sigma model.}.

How the dilaton influences dense matter depends on the mass of the dilaton that enters in the dynamics. At present the structure of the dilaton -- in fact the structure of low-energy scalars in general -- is not well understood~\cite{scalars}. In \cite{LR09}, two extreme cases were taken, a low mass object at $\sim 700$ MeV and a high mass at $\sim 3$ GeV, the two giving drastically different scenarios. Given that the ``hard" dilaton whose excitation could be of the order of the high mass object taken, the latter may not be relevant to the phenomenon concerned whereas the low mass object $f_0(600)$ is most likely to be relevant.

\begin{figure}[hbtp]
\centerline{\epsfig{file=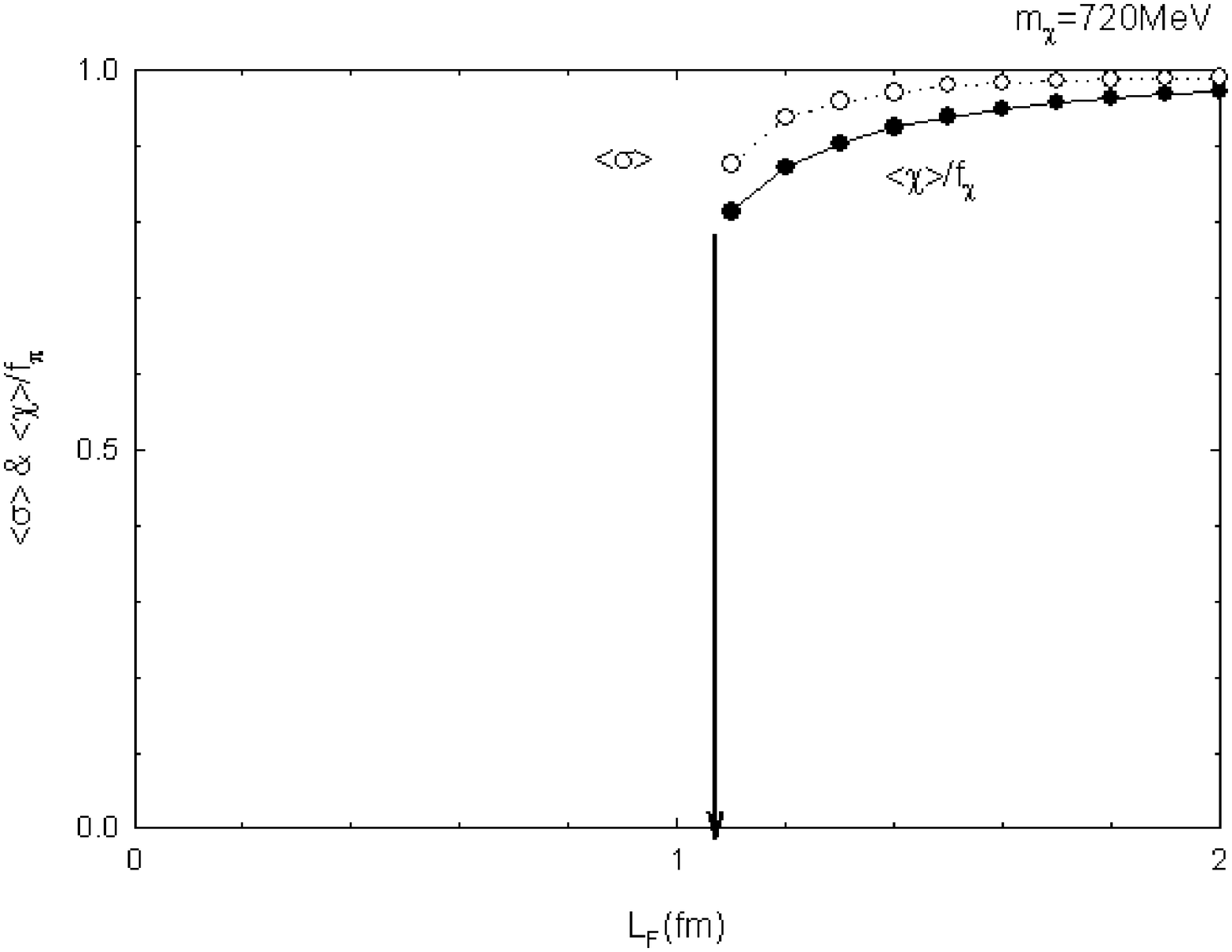,width=6.0cm,angle=0}
\epsfig{file=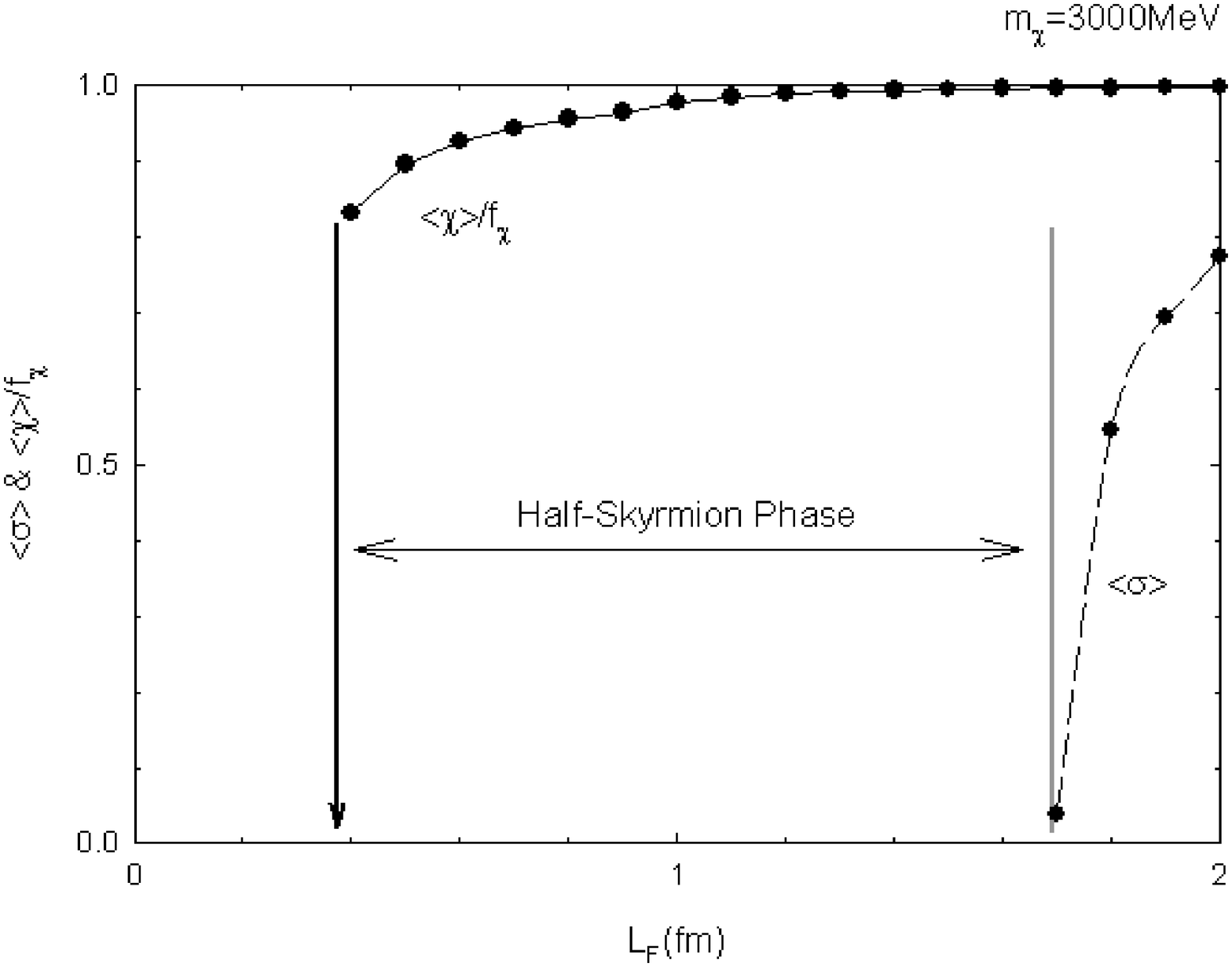,width=6.0cm,angle=0}}
\caption{Behavior of
$\langle\chi\rangle$ and $\langle\sigma\rangle\propto \la\bar{q}q\ra$ where $\sigma=\frac 12 {\rm Tr}U$ as a function
of lattice size for ``light" dilaton mass $m_\chi=720$ MeV (left figure) and for ``heavy" dilaton mass
$m_\chi=3000$ MeV (right figure).}
\label{half-skyrmion}
\end{figure}

The result of the Lagrangian (\ref{lagtot}) put on an FCC crystal~\cite{PRV08} is shown in Fig.~\ref{half-skyrmion}. For a heavy dilaton with mass $m_\chi \gg 1$ GeV, there is a distinctive phase in which $\la\bar{q}q\ra^*\propto \la {\rm Tr} U\ra^*=0$ but $f_\pi^*\sim \la\chi\ra^*\neq 0$. This phase has the skyrmions fractionized into half-skyrmions. However if the dilaton is light, say, $m_\chi\sim 700$ MeV, the 1/2-skyrmion phase shrinks to a point. The model cannot describe the confinement-deconfinement transition but one expects that above the transition point, there could be a deconfined quark phase or a color superconducting phase.  In either case, the pion decay constant $f_\pi^*$ and the $\omega$ mass $m_\omega^*$ are predicted to decrease as density increases, both going to zero at the critical point (in the chiral limit).

The important role of the dilaton in the presence of the $\omega$ meson can be seen by modifying the homogeneous Wess-Zumino term (\ref{fwzterm}). As mentioned, this is a special form gotten in certain approximations but its property is expected to be generic. Suppose that one sets $\kappa=1$ in (\ref{fwzterm}) and take
\bea
{\cal L}_{hWZ}^\prime &=& \textstyle\frac{3}{2} g \omega_\mu B^\mu.\label{fwztermp}
\eea
In fact, one naively expects this to be of the correct form from the point of view of the scaling dimension of the hWZ term which is 4 if $\kappa=1$. However as argued in \cite{LR09}, with the two dilatons $\chi_{s,h}$, one can construct (\ref{fwzterm}) -- with the exponent 3 on $\kappa$ -- valid for the soft-dilaton sector without violating the scale invariance. Now what happens with the skyrmion matter simulated on FCC crystal using the hWZ Lagrangian (\ref{fwztermp}) is a disastrous result totally at odds with nature~\cite{PRV-vector}: Both $f_\pi^*$ and $m_\omega^*$ are found to {\em increase} with increasing density rather than decrease as desired.

This feature is not difficult to understand. The key point is that the $\omega$ meson gives rise to a Coulomb potential. The hWZ term then leads to the repulsive interaction, contributing to the energy per baryon, $E/B$, of the form
\bea
(E/B)_{hWZ}
 = \frac{9g^2}{16}\textstyle \displaystyle
\int_{\mbox{\scriptsize Box}} d^3x \int d^3 x^\prime
B_0(\vec{x}) \frac{e^{-m_\omega^*}
|\vec{x}-\vec{x}^\prime|} {4\pi|\vec{x}-\vec{x}^\prime|}
B_0(\vec{x}^\prime). \label{omega}
\eea
What is important is that this repulsive interaction turns out to dominate over other terms as density increases. Now while the integral over $\vec{x}$ is defined in a single lattice (FCC) cell, that over $\vec{x}^\prime$ is not, so will lead to a divergence unless tamed. In order to prevent the $(E/B)_{hWZ}$ from diverging, $m_\omega^*$ has to {\em increase} sufficiently fast. And since $m_\omega^*\sim f_\pi^* g$ in this HLS model, for a fixed $g$, $f_\pi^*$ must therefore {\em increase}. In fact this phenomenon is a generic feature associated with the role that the vector mesons in the $\omega$ channel play in dense medium. This feature, however, is at variance with nature: QCD predicts that the pion decay constant tied to the chiral condensate should {\em decrease} and go to zero (in the chiral limit) at the chiral transition.

The suppression by the soft dilaton of the repulsion at high density has an important consequence on the maximum stable mass of neutron stars as described below.

\section{Vector Symmetry at High Density}
 What is perhaps the most significant for dense matter near chiral restoration is that the 1/2-skyrmion (or pseudogap) state exhibits an emerging or ``enhanced" symmetry. In HLS$_1$ theory, the 1/2-skyrmion state has the chiral $SU(N_f)_L\times SU(N_f)_R$ symmetry restored. Thus  for $n_{1/2S}\leq n < n_c^{\chi SR}$,
 \bea
 (F_\sigma/F_\pi)^2\equiv a=1, \ \  F_\pi\neq 0,
 \eea
 which corresponds to $\gamma=0$ in Eq.~(\ref{hls1}). Note however that the gauge coupling $g\neq 0$, so the vector meson remains  massive. Since the vector meson  is massive, $F_\sigma$ is the decay constant for the longitudinal component of the vector meson, not of a free scalar. The gauge coupling $g$ goes to zero, however, at chiral restoration, $n=n_c^{\chi SR}$. This corresponds to Georgi's ``vector limit."  As noted by Georgi~\cite{georgi}, at this point the symmetry ``swells" to $SU(N_f)^4$, with $\xi_L$ and $\xi_R$ transforming under independent $SU(N_f)\times SU(N_f)$ symmetries,
 \bea
 \xi_L\rightarrow h_L(x) \xi_L L^\dagger, \ \  \xi_R\rightarrow h_R(x) \xi_R R^\dagger,
 \eea
 where ${L,R}$ and $h_{L,R}$ are the unitary matrices generating the corresponding global and local $SU(N_f)$ groups. The hidden local symmetry is the diagonal sum of $SU(N_f)_{h_L}$ and $SU(N_f)_{h_R}$. Away from the vector limit, the non-zero gauge couplings break the vector symmetry explicitly producing the nonzero vector meson mass and couplings for the transverse components of the vector mesons. In terms of this symmetry pattern, we see that the pseudogap phase is the regime where one has $a=1$ ($\gamma=0$) and the gauge coupling $g$ is weak but non-zero.

 It is noteworthy that while chiral symmetry is restored, the quarks are confined in hadrons. This suggests to identify the hadronic freedom (or pseudogap) regime to be ``quarkyonic" as predicted in large $N_c$ QCD.

\subsection{Hadronic freedom}
The matter between the 1/2-skyrmion threshold density $n_{1/2S}$ and $n_{\chi SR}$ with the gauge coupling $g\rightarrow 0$ has been referred to as ``hadronic freedom" region with $n_{1/2S}$ identified as a ``flash density" $n_{flash}$ in analogy to the ``flash temperature" in hot medium as defined below. This pseudogap region has an important astrophysical implication. With $\gamma\rightarrow 0$
($a\rightarrow 1$), the gauge coupling $g$ goes to zero as density approaches $n_c^{\chi SR}$, so hadrons interact weakly in that regime. In \cite{BLPR-kcond}, this reasoning was used to predict kaon condensation at a density $\sim 3
n_0$. There the assumption was that kaons must condense somewhere
between the flash density $n_{flash}$ and $n_c^{\chi SR}$. Therefore one can start from the vector
manifestation fixed point of HLS theory with $a= 1$ and $g=0$ but
$F_\pi\neq 0$. This calculation reinforced the previous conclusion
that kaons must condense {\em before} any other phase changes can
take place and hence determine the fate of compact stars. This is
reviewed in \cite{BLR07}.

In relativistic heavy-ion collisions, one is dealing with high temperature and relatively low density. There is no equivalent hadronic description comparable to dense skyrmions for what happens between the chiral transition temperature $T_{\chi SR}$ and the ``flash temperature" $T_{flash}\sim 120$ MeV at which hadrons, in particular, the $\rho$ meson, go  nearly on-shell. There are however lattice calculations with dynamical quarks~\cite{miller} on thermodynamic properties of hot matter which indicate that at a temperature corresponding to $T_{flash}$ at which the condensate of the soft dilaton $\la\chi_s\ra^*$ (the asterisk here denotes temperature dependence) starts ``melting," vanishing at the chiral transition temperature $T_{\chi SR}\sim 200$ MeV. Between $T_{flash}$ and $T_{\chi SR}$, the gauge coupling must be small according to the hidden local symmetry in the vector manifestation~\cite{HY:PR}. Since temperature induces violation of the vector dominance in the photon-pion coupling, $a$ must be approaching 1. Thus with $g\approx 0$ and $a\approx 1$, this region can be considered as the temperature counter part of the ``hadronic freedom" established above in density. In \cite{BHHRS}, it has been proposed that dileptons decouple from the $\rho$ meson in this hadronic freedom region, which could explain the recent dilepton measurements at CERN and RHIC where no evidences for precursors to chiral symmetry restoration are seen.

 \begin{figure}[htb]
\centerline{\epsfig{file=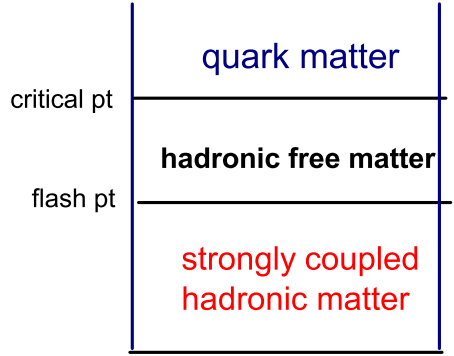,width=7.0cm,angle=0}}
\caption{A schematic picture of the proposed phase diagram modified by the
hidden local symmetry in the vector manifestation that is
conjectured to lead to ``hadronic freedom."  The ``critical (flash) pt" corresponds to $n_{\chi SR}$ ($n_{flash}$) in density and $T_{\chi SR}$ ($T_{flash}$) in temperature.} \label{phasediagram}
\end{figure}

In sum, a new hitherto unsuspected phase -- that is drastically different from the currently accepted one -- emerges from the above observations, i.e., the ``hadronic freedom phase," that connects a possible quark gluon phase to the normal hadronic phase, that ranges in the $(T,n)$ plane from $T_{flash}\lsim T\lsim T_{\chi SR}$ and $n=0$ to $n_{flash}\lsim n\lsim n_{\chi SR}$ and $T=0$ -- which may be identified with the quarkyonic matter at large $N_c$. A schematic form of the new phase structure is given in Fig.~\ref{phasediagram}. Here the ``quark matter" stands for what might be identified with ``sQGP"~\footnote{The state just above $T_{\chi SR}$ (and at low density) is not understood. In fact, it may have nothing to do with sQGP. All one can say at present is that it is most likely in the Wigner phase.} at $T\gsim T_{\chi SR}$ and $n\approx 0$ and mnemonically for the variety of color superconducting states with chiral symmetry either restored or broken or true quark matter with chiral symmetry restored. The case where the CFL phase figures there presents an interesting, though academic, case as discussed below.

\subsection{The fate of neutron stars}
The suppression by the dilaton $\chi_s$ of the repulsion brought about by the $\omega$ exchange between baryons has an interesting consequence on the fate of compact stars that are more massive than some critical mass $M_{crit}$ of the star. Phrased in terms of an effective field theory (EFT) for nuclear matter, the taming of the $\omega$ repulsion in (\ref{omega}) can be understood as follows. In nuclear EFT, (\ref{omega}) represents the contribution to $(E/B)$ from the mean field of a four-Fermi interaction in the effective Lagrangian of the form
\bea
\frac 12 {C_\omega^*}^2 (\overline{N}\gamma_\mu N)^2\label{fourfermi}
\eea
where $N$ is the nucleon field and ${C_\omega^*}^2$ is a constant proportional to $g^2/m_\omega^2$ coming from the $\omega$ exchange between two nucleons. It was observed~\cite{PRV-vector} that if in (\ref{omega}), the coupling constant $g$ is taken to be a constant, the repulsion grows in the skyrmion matter as density increases. This is translated in (\ref{fourfermi}) as the coefficient ${C_\omega^*}^2$ growing with density. This means that the vector meson mass $m_\omega$ is decreasing at increasing density with the vector coupling held constant. The remedy to this disease discussed above and given in the references \cite{PRV08,LR09} correspond to the vector coupling decreasing in some proportion to the mass $m_\omega$. In fact, in HLS$_1$ theory, the coupling constant $g$ is to drop proportionally to the quark condensate, and this circumvents the necessity for the $\omega$ mass to increase to counter the increasing repulsion. The intrinsic density dependence of the gauge coupling constant in HLS$_1$, when truncated at the four-Fermi interaction level, subsumes, among others, three- and more-body forces and hence the repulsion that is generated when $g$ is held constant can be considered as an effect of repulsive many-body forces. This is indeed what is found in specific model studies in many-body nuclear physics approaches~\cite{panda} where the many-body repulsion would lead to the maximum neutron star mass $\gsim 2M_\odot$ while it leaves unaffected the equation of state at the equilibrium density of normal nuclear matter. Such a repulsion sourced by many-body forces, if unsuppressed, would push kaon condensation to a density $n\gsim 7n_0$~\cite{panda2}, so that the maximum stable neutron star mass $M_{max}\simeq 1.56M_\odot$ conjectured by Brown and Bethe~\cite{brown-bethe,BLR07} would be ruled out. Thus the role of the light dilaton which renders the skyrmion matter consistent with HLS$_1$ theory is found to be crucial for the physics of compact stars. This issue will be addressed elsewhere.

\section{Transition from Nuclear Matter to CFL Phase}
At asymptotically high density, diquarks will condense to a form where color and flavor get locked. Here chiral symmetry is again spontaneously broken by the diquark condensate. Although it is not clear whether the color-flavor locked state is relevant for compact stars, so this phenomenon could be purely academic, it is nonetheless a theoretically interesting object. Now it is tempting to think of the phase transition from normal baryonic matter to quark matter going via the 1/2-skyrmion phase at some high density in analogy to the condensed matter case. To see whether this analogy can be made closer, let us consider the CFL phase of quark matter. In the real world of two (u and d) light flavors and one heavy (s) flavor,  a variety of model calculations predict a multitude of superconducting states, some unstable and some others (such as LOFF crystalline) presumably stable, but we are going to consider, for simplicity, the CFL configuration which is favored for degenerate quark masses. Furthermore, there is also a possibility that the CFL phase can come down in density all the way to the nuclear matter density for a light-enough s-quark mass, say, in the presence of strong $U(1)_A$ anomaly~\cite{CFL-anomaly}.

 Since in the CFL phase, the global color symmetry $SU(3)$ is completely broken and chiral $SU(3)_L\times SU(3)_R$ (for $N_f=3$) is broken down to the diagonal subgroup $SU(3)_V$~\cite{cfl-rev}, low-energy excitations can be described by the coordinates $\xi_{C+L}\in SU(3)_{C+L}$ and $\xi_{C+R}\in SU(3)_{C+R}$ given in terms of the octet pseudoscalar $\pi$ and  the octet scalar $s$. The scalars are eaten up by the gluons which become massive and are mapped one-to-one to the vector mesons present in the hadronic sector. The Lagrangian that describes low-energy excitations is of the same local gauge invariant form as (\ref{hls1}). The gauge symmetry here is explicit, not hidden as in the hadronic sector  but we will nonetheless call it HLS'. Now as in the hadronic sector, the HLS' Lagrangian supports solitons that carry fermion number $B$, which are nothing but skyrmions~\cite{qualiton}. The CFL soliton is called ``superqualiton" to be distinguished from the soliton in the hadronic phase.  It is actually a quark excitation on top of the vacuum with condensed Cooper pairs, effectively color singlet with spin 1/2. But in this formulation, it is a topological object.

 Given the skyrmion matter for $n\lsim n_{1/2S}$ and the superqualiton matter for $n\gsim n_c^{\chi SR}$, the transition from nuclear matter to the CFL matter can be considered as a skyrmion-superqualiton transition with half skyrmions figuring in between. In both the skyrmion phase and the superqualiton phase, chiral symmetry is spontaneously broken and quarks are confined. The order parameters are however different, the former with $\la\bar{q}q\ra$ and the latter with $\la qq\ra$. The 1/2-skyrmion phase sandwiched by the two Nambu-Goldstone phases has chiral symmetry restored but quarks are still confined inside hadrons. Therefore the conjectured phase change takes place in a confined phase.  This is the analogy to the N\'eel-VBS transition with half-skyrmions (spinons) at the boundary~\cite{senthiletal}.
 Independently of whether this analogy is just a coincidence or has a non-trivial meaning, what is significant is that the pseudogap region can deviate strongly from the Fermi-liquid state that is usually assumed in studying color superconductivity.

\section{Further Remarks} The main assumption made in this note is that the dense
skyrmion matter simulated in a crystal using HLS$_1$ Lagrangian represents dense baryonic matter. There are several questions one can raise here.

The first is whether there are no other crystal configurations that could (1) give a lower ground state and (2) induce different skyrmion fractionization. The answer to this is not known. It is unquestionably an important question to address. For instance,  in (2+1) dimensions, while a baby-skyrmion fractionizes into two half-skyrmions for the known square-cell configuration, it is the hexagonal configuration that has the minimal energy and induces the fractionization of a baby-skyrmion into four {\em quarter-skyrmions}~\cite{karliner}.

Given that the skyrmion-1/2skyrmion transition scenario is anchored on the crystalline structure at the mean field level, one wonders whether quantum fluctuations would not wash out the soliton structure of the 1/2-skyrmion matter. As mentioned in \cite{Skyrme-matter}, since nuclear matter is known to be a liquid, not a crystal, it might be that quantum fluctuations would ``melt" the crystal. The phase change could then  be merely a lattice  artifact although at high density baryonic matter is favored to be in the form of a crystal. In addition,  the spin and statistics of the 1/2-skyrmion would require quantization. It seems highly plausible however that given that the transition involved here is a topology change, the phase change be robust against quantum fluctuations. Similar issues are raised in condensed matter physics where the concept of ``topological order" is invoked for robustness of topology-changing phase transitions.

It should be stressed that the half-skyrmions ``live" in the confined phase, i.e., hadrons, so need not have to be identified with the QCD degrees of freedom, i.e., the quarks with color and fractional electric charges.

The next unanswered question is the mechanism for the
fractionization of a skyrmion to two half-skyrmions at $n_{1/2S}$. The
fractionization under certain external conditions seems generic, taking place both in
(2+1) and (3+1) dimensions. The treatment made in this note was
based on energetics considerations but the mechanism was left
unclarified. In the condensed matter case discussed in
\cite{senthiletal}, the key role for the fractionization is played
by the emergent $U(1)$ gauge field and its monopole structure.  The
pair of half-skyrmions (referred to as ``up-meron" and
``down-anti-meron" in \cite{senthiletal}) are confined -- or bound -- to a single skyrmion
in both the initial N\'eel state and the final VBS state but the
skyrmion fractionizes into half skyrmions at the boundary due to the
``irrelevance" of the monopole tunneling, with an emergent global symmetry not present in the many-body Hamiltonian. It would be exciting to
see  a similar mechanism at work in the present case. It could
elucidate what the hadronic phase could be at the doorway to color
superconductivity, should the latter survive the black-hole collapse following
kaon condensation~\cite{BLR07}. In this regard, it would be
interesting to investigate the skyrmion-1/2skyrmion transition in
terms of the instantons and merons of  5D Yang-Mills Lagrangian of
hQCD which would reveal the role of the infinite tower.

If the pseudogap phase is indeed the hadronic freedom region, how
can one exploit the background provided by the half-skyrmion matter
for describing kaon condensation? One may embed and bind $K^{-}$'s in  dense
half-skyrmion matter where $a\approx 1$ and $g\sim 0$ and exploit
that  in compact stars, electrons with high chemical potential decay
into $K^-$'s once the kaon mass falls sufficiently low and the kaons
Bose-condense. To do this calculation, it may be necessary to know
what the quantum structure of the half-skyrmion phase is.
\subsection*{Acknowledgments}
This work was supported in part by the WCU project of Korean Ministry of Education, Science and Technology (R33-2008-000-10087-0)

\bibliographystyle{ws-rv-van}

\begin{thebibliography}{99}

\bibitem{cfl-rev} For review, see K. Rajagopal and F. Wilczek, {\it At
the frontier of particle physics: Handbook of QCD}\ ed by M. Shifman
(World Scientific, Singapore, 2001) Vol. 3, p.2061.

\bi{BLR07} For review, see G.E. Brown, C.-H. Lee and M. Rho, ``Recent developments on kaon condensation and its astrophysical implications," {\it Phys.\ Rept.}\  {\bf 462} (2008) 1 [arXiv:0708.3137 [hep-ph]].

\bi{MR-halfskyrmion} M.~Rho, ``Hidden local symmetry and dense half-skyrmion matter,''
  arXiv:0711.3895 [nucl-th].


\bi{georgi} H. Georgi, ``New realization of chiral symmetry,"  {\it Phys.\ Rev.\ Lett.}\ {\bf 63} (1989) 1917;  ``Vector realization of chiral symmetry,"
{\it Nucl.\ Phys.} {\bf B331} (1990) 311.

\bibitem{HY:PR} M. Harada and K. Yamawaki, ``Hidden local symmetry at loop: A new perspective of composite
gauge bosons and chiral phase transition,"   {\it Phys. Rept.}\ {\bf 381} (2003) 1.


\bibitem{quarkyonic} L.~McLerran and R.~D.~Pisarski,
  ``Phases of cold, dense quarks at large $N_c$,''
  {\it Nucl.\ Phys.}\ {\bf A796} (2007) 83.

\bibitem{hsu} S.~D.~H.~Hsu, F.~Sannino and M.~Schwetz,
  ``Anomaly matching in gauge theories at finite matter density,''
  {\it Mod.\ Phys.\ Lett.}\  A {\bf 16} (2001) 1871
  [arXiv:hep-ph/0006059].


\bibitem{BHHRS}  G.~E.~Brown, M.~Harada, J.~W.~Holt, M.~Rho and C.~Sasaki,
  ``Hidden local field theory and dileptons in relativistic heavy ion collisions,'' 
{\it Prog.\ Theor.\ Phys.},\ in press, arXiv:0901.1513 [hep-ph].



\bibitem{senthiletal} T. Senthil et al., ``Deconfined quantum critical points," {\it Nature} {\bf 303} (2004) 1490.

\bi{MR-Yukawa} M. Rho,  ``Hidden local symmetry and the vector
manifestation of chiral symmetry in hot and/or dense matter,"
{\it Prog. Theor. Phys. Suppl.}  {\bf 168}(2007) 519.

\bi{battye-sutcliffe} R. Battye and P. Sutcliffe,
{\it\pr}\ {\bf C73} (2006) 055205; R. Battye et al, hep-th/0605284.

\bibitem{PRV08} B.-Y. Park, M. Rho and V. Vento,  ``The role of the dilaton in  dense skyrmion matter", {\it Nucl.\ Phys.}\  {\bf A807} (2008) 28.

\bibitem{LR09}   H.~K.~Lee and M.~Rho,
  ``Dilatons in hidden local symmetry for hadrons in dense matter,''
  arXiv:0902.3361 [hep-ph].

\bibitem{SSmodel} T.~Sakai and S.~Sugimoto,  ``Low energy hadron physics in holographic QCD,''
  {\it Prog.\ Theor.\ Phys.}\  {\bf 113} (2005) 843; 
  ``More on a holographic dual of QCD,''
  {\it Prog.\ Theor.\ Phys.}\  {\bf 114} (2005) 1083.
 

\bi{HRYY} D.-K. Hong, M. Rho, H.-U. Yee and P. Yi,  ``Chiral dynamics of baryons from string theory," {\it Phys. Rev.}\  {\bf D76} (2007) 061901;  ``Dynamics of baryons from string theory and vector dominance,"  {\it JHEP} {\bf 09}, 063 (2007); ``Nucleon form factors and hidden symmetry in holographic QCD," arXiv:0710.4615 [hep-ph].

\bibitem{SS} K.~Hashimoto, T.~Sakai and S.~Sugimoto,
  ``Holographic baryons : Static properties and form factors from gauge/string
  duality,''
  arXiv:0806.3122 [hep-th];  H.~Hata, T.~Sakai, S.~Sugimoto and S.~Yamato,
  ``Baryons from instantons in holographic QCD,''
  arXiv:hep-th/0701280.


\bibitem{Kim-Zahed}
  K.~Y.~Kim and I.~Zahed,
  ``Electromagnetic baryon form factors from holographic QCD,''
  {\it JHEP} {\bf 0809} (2008) 007
  [arXiv:0807.0033 [hep-th]].

\bibitem{emergent} For a lucid exposition, see N. Arkani-Hamed, H. Georgi
and M.D. Schwartz,  ``Effective field theory for massive
gravitons and gravity in theory space," {\it Ann. Phys.} {\bf 305} (2003) 96.


\bibitem{son-stephanov} D.T.  Son and M.A. Stephanov, `` QCD and dimensional deconstruction,"  {\it\pr}\
{\bf D69} (2004) 065020.

\bi{HMY} M. Harada, S. Matsuzaki and K. Yamawaki, `` Implications of holographic QCD in ChPT with hidden local symmetry,"  {\it \pr}\ {\bf D74} (2006) 076004.


\bi{BR91} G.E.  Brown and M. Rho, ``Scaling effective Lagrangians in a dense medium,"  {\it Phys. Rev. Lett.} {\bf 66} (1991) 2720.

\bi{Skyrme-matter}  B.-Y. Park, D.-P. Min, M. Rho and V. Vento, \np\ {\bf A707} (2002) 381;  H.-J. Lee, B.-Y. Park, D.-P. Min, M. Rho and V. Vento, {\it\np}\ {\bf A723} (2003) 427.

\bi{klebanov} I.R. Klebanov,  ``Nuclear matter in the Skyrme model," {\it Nucl. Phys.} {\bf B262} (1985) 133.

\bi{FCC} M. Kugler and S. Shtrikman,  ``Skyrmion crystals and their symmetries,"   {\it Phys. Rev.}\ {\bf D40} (1989) 3421;  L. Castillejo, P.S.J.  Jones, A.D. Jackson and J.J.M  Verbaarschot,  ``Dense skyrmion systems," {\it Nucl. Phys.}\ {\bf A501} (1989) 801.

\bi{karliner} I. Hen and M. Karliner, ``Hexagonal structure of baby skyrmion lattice," arXiv:0711.2387 [hep-th].


\bibitem{BHLR06} G.E. Brown, J.W.  Holt, C-H. Lee and  M. Rho, ``Late hadronization and matter formed
at RHIC: Vector manifestation, BR scaling and hadronic freedom," {\it Phys. Rept.} {\bf 439} (2006) 161.

\bi{goldhaber-manton} A.S. Goldhaber and N.S. Manton, `` Maximal symmetry of the Skyrme crystal,"  {\it \pl}  {\bf B198} (1987) 231.

\bi{zarembo} K. Zarembo, ``Possible pseudogap phase in QCD,"  {\it JETP Lett.}\ {\bf 75} (2002) 59.

\bi{scalars} For a recent discussion, see  A.~H.~Fariborz, R.~Jora and J.~Schechter,
  ``Global aspects of the scalar meson puzzle,''   arXiv:0902.2825 [hep-ph].


\bi{PRV-vector} B.-Y. Park, M. Rho and V. Vento, ``Vector mesons and dense skyrmion matter,"  {\it Nucl. Phys.}\ {\bf A736} (2004)129.

\bibitem{miller} D.E. Miller, ``Lattice QCD
 calculations for the physical equation of state," {\it Phys. Rept.}\ {\bf 443} (2007) 55

\bi{panda}  A.~Akmal, V.~R.~Pandharipande and D.~G.~Ravenhall,
  ``The equation of state for nucleon matter and neutron star structure,''
  {\it Phys.\ Rev.}\  C {\bf 58} (1998) 1804
  [arXiv:nucl-th/9804027].

\bi{panda2}  V.~R.~Pandharipande, C.~J.~Pethick and V.~Thorsson,
  ``Kaon energies in dense matter,'' {\it Phys.\ Rev.\ Lett.}\  {\bf 75} (1995) 4567
  [arXiv:nucl-th/9507023].


\bi{brown-bethe} G.E. Brown and H.A. Bethe,
``A scenario for a large number of low-mass black holes in
the galaxy," {\it Astrophys. J.} {\bf 423} (1994) 659.







\bi{OMPR07}  Y. Oh, B.-Y. Park and M. Rho, work in progress.

\bi{CFL-anomaly} T. Hatsuda,  M.  Tachibana, N. Yamamoto and G. Baym, ``New critical point induced by the axial anomaly in dense QCD, " {\it \prl}\ {\bf 97} (2006) 122001.

\bi{qualiton}  D.K  Hong, M.  Rho and I.  Zahed,  ``Qualitons at high density," {\it Phys. Lett.}\ {\bf B468} (1999) 261.


\bi{BLR-STAR}   G.E. Brown, C-H. Lee and  M. Rho, ``Vector manifestation of hidden local symmetry, hadronic freedom, and the STAR $\rho^0 /\pi^-$  ratio,"
{\it Phys. Rev.}\ {\bf C74} (2006) 024906.

\bi{BLPR-kcond} G.E. Brown, C-H. Lee, H.-J. Park and  M. Rho,  ``Study of strangeness condensation by expanding about the fixed point of the Harada-Yamawaki vector manifestation,"  {\it Phys. Rev. Lett.}\ {\bf 96} (2006) 062303.


\end{thebibliography}

\end{document}